\documentclass{internoise2026}

\begin{document}

\title{Direction-preserving Spatial Audio Signal Enhancement in The Spherical Harmonic Domain}

\maketitle

\ifcase 3 %

\or %

  \begin{flushleft}
    GivenName FamilyName
    
    \institution[]{%
      mail@example.com\\
      Institution name\\
      Full address Line 1\\
      Full address Line 2
    }
  \end{flushleft}

\or %

  \begin{flushleft}
    GivenName FamilyName,
    GivenName FamilyName,\markpresenting{}
    GivenName FamilyName\markcorresponding{}
    
    \correspondingemail{mail1@example.com}
    \presentingemail{mail2@example.com}
    
    \institution[]{%
      Institution name\\
      Full address Line 1\\
      Full address Line 2
    }
  \end{flushleft}

\or %
  
  \begin{flushleft}
    Huawei Zhang,\markcorrespondingpresenting{1}
    Jihui Aimee Zhang,\markinstitution{2, 1}
    Huiyuan June Sun,\markinstitution{1}
    Prasanga Samarasinghe\markinstitution{1}
    
    \correspondingpresentingemail{u7178294@anu.edu.au}
    
    \begin{multicols}{2} %
    \institution[1]{%
      The Australian National University\\
      Canberra, Australia
    }
    \institution[2]{%
      The University of Queensland\\
      Brisbane, Australia
    }
    \end{multicols}
  \end{flushleft}

\fi

\smallskip
\begin{abstract}
Spatial audio signal enhancement (SASE) aims to separate a desired soundfield from interference and noise while preserving its spatial cues.
However, existing algorithms pay limited attention to preserving the spatial cues of the residual interference, which reduces their effectiveness in practical applications.
In this paper, we propose a direction-preserving SASE algorithm based on multi-output beamforming.
The proposed algorithm extracts the desired soundfield while preserving the spatial cues of both the desired soundfield and the residual interference soundfield.
The simulation study shows that the proposed algorithm achieves performance comparable to the existing SASE algorithm in terms of estimation error, while it better preserves the spatial cues of the residual interference within a region of interest, in a noisy and reverberant environment.

\end{abstract}

\section{Introduction}
\label{sec:introduction}
Spatial audio signal enhancement (SASE) has been developed in the past decade \cite{hu2020acoustic, lugasi2022spatial, borrelli2018denoising, herzog2020direction, hafsati2019sound, herzog2022ambisep, herzog2023ambisep, matsuda2023sound, zhang2025spatial, murakami2025directional}, which extracts a desired soundfield from the mixed microphone recording while preserving the desired spatial cues.
This technology plays a fundamental role in many spatial audio applications, including soundfield decomposition \cite{matsuda2023sound}, soundfield recording \cite{zhang2017surround}, soundfield reproduction \cite{spors2013spatial, zhang2017surround, yang2022review}, and spatial active noise control \cite{ma2020active, shi2023active, zhang2025spherical}.

With the development of SASE, there is a growing requirement to preserve the spatial cues of the residual interference.
This requirement arises for two main reasons:
(1) An immersive listening experience depends on preserving spatial cues not only for the desired soundfield but also for the residual interference.
(2) Speech intelligibility is influenced by the spatial cues of both the desired soundfield and the residual interference \cite{licklider1948influence, rennies2018benefit}.
Despite this, most existing algorithms do not adequately preserve the spatial cues of the residual interference, often introducing distortions.
These distortions can degrade spatial perception and negatively affect the overall listening experience.

To address this limitation, direction-preserving SASE algorithms have been proposed in \cite{herzog2020direction}.
However, these approaches either rely on impractical assumptions, such as prior knowledge of the power spectral density (PSD) matrix of the desired soundfield, or compromise the effectiveness of interference reduction \cite{cornelis2009theoretical}.
Related efforts in binaural beamforming \cite{hadad2015theoretical} pursue a similar objective without these limitations by introducing spatial constraints derived from relative transfer functions.
Nevertheless, these direction-preserving binaural beamforming algorithms are designed for two specific target points rather than a region of interest (ROI), which limits their applicability in spatial audio.

Relative Harmonic Coefficients (ReHCs) \cite{hu2020acoustic, hu2020semi} have been used for SASE \cite{hu2020acoustic, herzog2020direction, zhang2025spatial}.
Similarly to \cite{hadad2015theoretical}, an SASE algorithm proposed in recent work \cite{zhang2025spatial} derives spatial constraints based on ReHCs, to preserve the desired soundfield with its spatial cues. 
However, the spatial cues of the residual interference are not well preserved in \cite{zhang2025spatial}, indicating that the potential of ReHCs has not been fully investigated.

In this paper, motivated by the SASE algorithm \cite{zhang2025spatial} and the direction-preserving binaural beamforming algorithm \cite{hadad2015theoretical}, we propose a direction-preserving SASE algorithm in the spherical harmonic (SH) domain.
The proposed algorithm captures the desired soundfield with its spatial cues, and preserves the spatial cues of the residual interference soundfield within a ROI.
The algorithm is formulated as a multi-output beamforming approach based on ReHCs \cite{hu2020acoustic}, without requiring prior knowledge of the PSD matrix of the desired soundfield.
Building on \cite{zhang2025spatial}, we further derive additional spatial constraints from ReHCs to preserve the spatial cues of the residual interference soundfield.
Simulation results demonstrate that the proposed algorithm preserves the spatial cues of both the desired soundfield and the residual interference, while effectively reducing interference and noise.

The remainder of this paper is organized as the followings:
The problem is formulated in Sec.~\ref{sec:problem}.
We present the proposed direction-preserving SASE algorithm in Sec.~\ref{sec:enhancement}.
The Simulation study is discussed in Sec.~\ref{sec:simulation}. 
Sec.~\ref{sec:conclusion} concludes the paper.

\section{Problem Formulation}
\label{sec:problem}

\begin{figure}[t]
\centering
\includegraphics[width=10 cm]{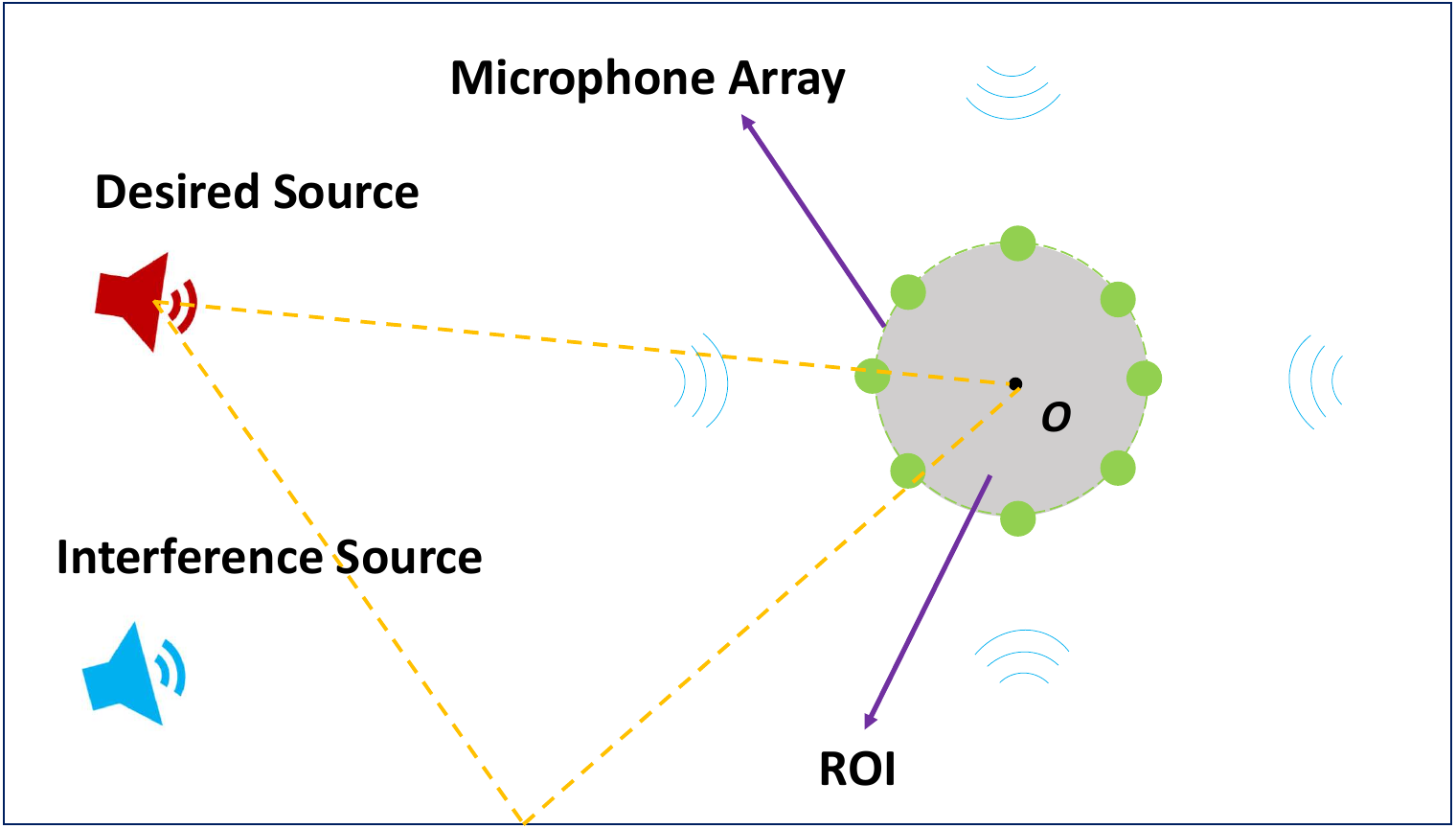}
\caption{The setup of the system, which consists of a desired source (red loudspeaker), an interference source (blue loudspeaker), a diffuse noise field (blue wave), and a spherical microphone array (green solid dots) at the origin $O$, the center of a spherical ROI (grey area). The orange dashed lines denote examples of acoustic paths from the desired source to the ROI.}
\label{fig:layout}
\end{figure}

As shown in Fig.~\ref{fig:layout}, there exists a desired source, an interference source, a diffuse noise field, and a spherical microphone array in a reverberant room, where a spherical ROI is located at $O$.
As a result, the acoustic environment contains a mixed soundfield composed of a desired soundfield, an interference soundfield, and the diffuse noise field.
The mixed soundfield $x(k, \mathbf{r})$ at an arbitrary point $\mathbf{r}$ can be decomposed into SH coefficients as \cite{rafaely2015fundamentals, williams1999fourier}
\begin{equation}
  \label{equ:SH_dec}
  x(k, \mathbf{r}) = \sum_{n=0}^{N_k} \sum_{m=-n}^n \tilde{x}_{nm}(k) j_n(kr) Y_{n m}(\theta, \phi),
\end{equation} 
where $\mathbf{r} \equiv \{r, \theta, \phi\}$, $r$, $\theta$, and $\phi$ are the radius, the elevation angle, and the azimuth angle, respectively, $k = 2 \pi f / c$ is the wavenumber, $f$ is the frequency, $c$ is the sound speed, $\tilde{x}_{nm}(k)$ is the respective SH coefficient of the $n$-th order and the $m$-th mode, $j_n(\cdot)$ is the Spherical Bessel function of the first kind, $Y_{nm}(\cdot)$ is the SH function, and $N_k$ is the maximum order of the soundfield for the wavenumber $k$ defined by \cite{rafaely2015fundamentals, williams1999fourier} 
\begin{equation}
  \label{equ:SHT order wavenumber}
  N_k = \left\lceil kR_0 \right\rceil = \left\lceil \frac{2 \pi f}{c}R_0 \right\rceil,
\end{equation} 
where $R_0$ represents the radii of both the microphone array and the ROI, and $\lceil \cdot \rceil$ is the ceiling function.

The measured mixed signals $\mathbf{x} = \{x(k, \mathbf{r})\}$ at the microphone array are denoted by 
\begin{equation}
  \label{equ:mic_divided}
  \begin{split}
    {\mathbf{x}} &=  {\mathbf{u}} + {\mathbf{v}} + {\mathbf{z}},
  \end{split}
\end{equation} 
where ${\mathbf{u} = \{u(k, \mathbf{r})\}}$, ${\mathbf{v} = \{v(k, \mathbf{r})\}}$, and ${\mathbf{z} = \{z(k, \mathbf{r})\}}$ denote the desired signals, the interference signals, and the diffuse noise signals, respectively.
After applying the Spherical Harmonic Transform (SHT) \cite{rafaely2015fundamentals, williams1999fourier} to the measured mixed signals $\mathbf{x}$, we obtain $L = (N_k+1)^2$ mixed SH coefficients $\tilde{\mathbf{x}} = \{\tilde{x}_{nm}(k)\}$, which can be decomposed as
\begin{equation}
  \label{equ:mixed_SH_divided}
  \begin{split}
    \tilde{\mathbf{x}} &= \tilde{\mathbf{u}} + \tilde{\mathbf{v}} + \tilde{\mathbf{z}},
  \end{split}
\end{equation} 
where $\tilde{\mathbf{u}}$, $\tilde{\mathbf{v}}$, and $\tilde{\mathbf{z}}$ are the SH coefficients representing the the desired soundfield, the interference soundfield, and the diffuse noise field, respectively.

We aim to obtain an estimate of desired SH coefficients, $\hat{\tilde{\mathbf{u}}} = \{ \hat{\tilde{{u}}}_{nm} \}$, from the mixed recording $\tilde{\mathbf{x}}$, which can be represented by
\begin{equation}
  \label{equ:estimated_SH_divided}
  \begin{split}
    \hat{\tilde{\mathbf{u}}} = \tilde{\mathbf{x}}_\text{res} = \tilde{\mathbf{u}}_\text{res} + \tilde{\mathbf{v}}_\text{res} + \tilde{\mathbf{z}}_\text{res},
  \end{split}
\end{equation} 
where $\tilde{\mathbf{x}}_\text{res}$, $\tilde{\mathbf{u}}_\text{res}$, $\tilde{\mathbf{v}}_\text{res}$, and $\tilde{\mathbf{z}}_\text{res}$ are the SH coefficients representing the residual mixed soundfield, the residual desired soundfield, the residual interference soundfield, and the residual diffuse noise field, respectively.
The objective of this work is to estimate $\hat{\tilde{\mathbf{u}}}$ such that it approximates the desired SH coefficients $\tilde{\mathbf{u}}$, while ensuring that $\tilde{\mathbf{v}}_\text{res}$ maintains spatial cues similar to those of the original interference SH coefficients $\tilde{\mathbf{v}}$.

\section{Direction-preserving spatial audio signal enhancement}
\label{sec:enhancement}
In this section, we propose a direction-preserving SASE algorithm based on ReHCs, which captures the desired soundfield with its spatial cues and preserve the spatial cues of the residual interference soundfield.

The proposed algorithm is formulated as a multi-output minimum variance distortionless response (MVDR) beamforming approach in the SH domain, consisting of $L$ beamformers $\tilde{\mathbf{b}} = \{\tilde{\mathbf{b}}_{nm}\}$ as a $L^2 \times 1$ vector \cite{zhang2025spatial, lugasi2022spatial, hadad2015theoretical, rafaely2015fundamentals}.
These beamformers $\tilde{\mathbf{b}}$ are required to extract the desired SH coefficients, reduce SH coefficients of interference and diffuse noise, and preserve the spatial cues of both the desired SH coefficients and the residual interference SH coefficients.

To satisfy the requirement, two sets of spatial constraints are introduced as follows.
First, a set of spatial constraints are derived from ReHCs of the desired soundfield \cite{zhang2025spatial}, represented by
\begin{equation}
  \label{equ:constraint_1}
  \begin{split}
    \tilde{\mathbf{b}}_{00}^H \tilde{\mathbf{h}}^u = \tilde{h}^u_{00}, \cdots, \tilde{\mathbf{b}}_{nm}^H \tilde{\mathbf{h}}^u = \tilde{h}^u_{nm}, \cdots, \tilde{\mathbf{b}}_{N_k N_k}^H \tilde{\mathbf{h}}^u = \tilde{h}^u_{N_k N_k},
  \end{split}
\end{equation}
where $\tilde{\mathbf{h}}^u = \{ \tilde{{h}}^u_{nm} \}$ represents ReHCs of the desired soundfield.
With these constraints defined in Eq.~(\ref{equ:constraint_1}), beamformers $\tilde{\mathbf{b}}$ are expected to extract the desired soundfield with its spatial cues.
Second, motivated by \cite{hadad2015theoretical}, we derive another set of spatial constraints from ReHCs of the interference soundfield.
The derived additional spatial constraints are represented by
\begin{equation}
  \label{equ:constraint_2}
  \begin{split}
    \frac{\tilde{\mathbf{b}}_{1-1}^H \tilde{\mathbf{h}}^v}{\tilde{\mathbf{b}}_{00}^H \tilde{\mathbf{h}}^v} = \frac{\tilde{h}^v_{1-1}}{\tilde{h}^v_{00}}, \cdots, \frac{\tilde{\mathbf{b}}_{nm}^H \tilde{\mathbf{h}}^v}{\tilde{\mathbf{b}}_{00}^H \tilde{\mathbf{h}}^v} = \frac{\tilde{h}^v_{nm}}{\tilde{h}^v_{00}}, \cdots, \frac{\tilde{\mathbf{b}}_{N_k N_k}^H \tilde{\mathbf{h}}^v}{\tilde{\mathbf{b}}_{00}^H \tilde{\mathbf{h}}^v} = \frac{\tilde{h}^v_{N_K N_k}}{\tilde{h}^v_{00}},
  \end{split}
\end{equation}
where $\tilde{\mathbf{h}}^v = \{ \tilde{{h}}^v_{nm} \}$ represents ReHCs of the interference soundfield.
With these constraints defined in Eq.~(\ref{equ:constraint_2}), beamformers $\tilde{\mathbf{b}}$ are expected to preserve the spatial cues of the residual interference soundfield.
These ReHCs $\tilde{\mathbf{h}}^u$ and $\tilde{\mathbf{h}}^v$ can be obtained as prior knowledge, respectively, as \cite{hu2020semi, markovich2015performance}
\begin{equation}
    \label{equ:rehc_desired}
    \tilde{\mathbf{h}}^u
    \approx \frac{\hat{\tilde{\mathbf{R}}}_{u} \tilde{\mathbf{e}}_1}{\tilde{\mathbf{e}}_1^H \hat{\tilde{\mathbf{R}}}_{u} \tilde{\mathbf{e}}_1},
\end{equation}
\begin{equation}
    \label{equ:rehc_interference}
    \tilde{\mathbf{h}}^{v}
    \approx \frac{\hat{\tilde{\mathbf{R}}}_{v} \tilde{\mathbf{e}}_1}{\tilde{\mathbf{e}}_1^H \hat{\tilde{\mathbf{R}}}_{v} \tilde{\mathbf{e}}_1},
\end{equation}
where $\tilde{\mathbf{e}}_1 = [1 \quad \mathbf{0}_{1 \times (L-1)}]^T$, $\hat{\tilde{\mathbf{R}}}_{u}$ and $\hat{\tilde{\mathbf{R}}}_{v}$ are the estimated SH-domain PSD matrices obtained from received microphone signals when only the desired soundfield exists and when only the interference soundfield exists, respectively.
Note that ReHCs are independent to the sound signal emitted at the corresponding source, so it is convenient to calculate ReHCs in advance.
Motivated by \cite{hadad2015theoretical, zhang2025spatial}, we develop a cost function with two sets of spatial constraints above by
\begin{equation}
  \label{equ:cost_singl_sh}
  \begin{split}
    \mathop{\operatorname{minimize}}\limits_{\tilde{\mathbf{b}}} \quad &\tilde{\mathbf{b}}_{00}^H \tilde{\mathbf{R}}_{v+z} \tilde{\mathbf{b}}_{00} + \cdots +\tilde{\mathbf{b}}_{nm}^H \tilde{\mathbf{R}}_{v+z} \tilde{\mathbf{b}}_{nm} + \cdots + \tilde{\mathbf{b}}_{N_kN_k}^H \tilde{\mathbf{R}}_{v+z} \tilde{\mathbf{b}}_{N_kN_k}\\
    \operatorname{subject \ to} \quad &\tilde{\mathbf{b}}_{00}^H \tilde{\mathbf{h}}^u = \tilde{h}^u_{00}, \cdots, \tilde{\mathbf{b}}_{nm}^H \tilde{\mathbf{h}}^u = \tilde{h}^u_{nm}, \cdots, \tilde{\mathbf{b}}_{N_k N_k}^H \tilde{\mathbf{h}}^u = \tilde{h}^u_{N_k N_k},\\
    \operatorname{subject \ to} \quad &\frac{\tilde{\mathbf{b}}_{1-1}^H \tilde{\mathbf{h}}^v}{\tilde{\mathbf{b}}_{00}^H \tilde{\mathbf{h}}^v} = \frac{\tilde{h}^v_{1-1}}{\tilde{h}^v_{00}}, \cdots, \frac{\tilde{\mathbf{b}}_{nm}^H \tilde{\mathbf{h}}^v}{\tilde{\mathbf{b}}_{00}^H \tilde{\mathbf{h}}^v} = \frac{\tilde{h}^v_{nm}}{\tilde{h}^v_{00}}, \cdots, \frac{\tilde{\mathbf{b}}_{N_k N_k}^H \tilde{\mathbf{h}}^v}{\tilde{\mathbf{b}}_{00}^H \tilde{\mathbf{h}}^v} = \frac{\tilde{h}^v_{N_K N_k}}{\tilde{h}^v_{00}},\\
  \end{split}
\end{equation}
where $\tilde{\mathbf{R}}_{v+z}$ is the SH-domain PSD matrices of interference and diffuse noise.
Based on the cost function in Eq.~(\ref{equ:cost_singl_sh}), we can design beamformers $\tilde{\mathbf{b}}$ to extract the desired soundfield, reduce the interference soundfield and diffuse noise, and preserve the spatial cues of both the extracted desired soundfield and the residual interference soundfield.

Further, we rewrite the cost function in Eq.~(\ref{equ:cost_singl_sh}) by
\begin{equation}
  \label{equ:cost_multi_sh_jrc}
  \begin{split}
    \mathop{\operatorname{minimize}}\limits_{\tilde{\mathbf{b}}} \quad &\tilde{\mathbf{b}}^H \tilde{\bm{\mathcal{R}}} \tilde{\mathbf{b}}\\
    \operatorname{subject \ to} \quad & \tilde{\mathbf{C}}^H \tilde{\mathbf{b}} = \tilde{\mathbf{a}},
  \end{split}
\end{equation}
where $\tilde{\mathbf{a}}$ is a $(2L-1) \times 1$ column vector denoted as
\begin{equation}
  \tilde{\mathbf{a}} = 
  \begin{bmatrix}
    \mathrm{conj}(\tilde{\mathbf{h}}^u) \\
    \mathbf{0}_{(L-1) \times 1}
  \end{bmatrix}_{(2L-1) \times 1},
\end{equation}
$\mathrm{conj}(\quad)$ represents the conjugate operation, $\tilde{\bm{{\mathcal{R}}}}$ and $\tilde{\mathbf{C}}$ are two large matrices, respectively, denoted as
\begin{equation}
  \tilde{\bm{{\mathcal{R}}}} = 
  \begin{bmatrix}
    \tilde{\mathbf{R}}_{v+z} & & & \\
     & \tilde{\mathbf{R}}_{v+z} & & \\
     & & \cdots & \\
     & & & \tilde{\mathbf{R}}_{v+z}
  \end{bmatrix}_{L^2 \times L^2},
\end{equation}
\begin{equation}
  \tilde{\mathbf{C}} = 
  \begin{bmatrix}
   \tilde{\mathbf{C}}^d & \tilde{\mathbf{C}}^{v}
  \end{bmatrix}_{L^2 \times (2L-1)},
\end{equation}
\begin{equation}
  \tilde{\mathbf{C}}^d = 
  \begin{bmatrix}
    \tilde{\mathbf{h}}^u & & & \\
     & \tilde{\mathbf{h}}^u & & \\
     & & \cdots & \\
     & & & \tilde{\mathbf{h}}^u
  \end{bmatrix}_{L^2 \times L},
\end{equation}
\begin{equation}
  \tilde{\mathbf{C}}^{v} = 
  \begin{bmatrix}
     -\tilde{{h}}^v_{1-1}\tilde{\mathbf{h}}^v & -\tilde{{h}}^v_{10}\tilde{\mathbf{h}}^v & & -\tilde{{h}}^v_{NN}\tilde{\mathbf{h}}^v\\
     \tilde{\mathbf{h}}^v & 0 & & 0 \\
      & \tilde{\mathbf{h}}^v & & 0 \\
     & & \cdots & \\
     0& & & \tilde{\mathbf{h}}^v
  \end{bmatrix}_{L^2 \times (L-1)}.
\end{equation}
The optimal solution of Eq.~(\ref{equ:cost_multi_sh_jrc}) can be also obtained as \cite{van1988beamforming}
\begin{equation}
  \label{equ:solution_sh_jrc}
  \tilde{\mathbf{b}} = \tilde{\bm{{\mathcal{R}}}}^{-1} \tilde{\mathbf{C}} [\tilde{\mathbf{C}}^H \tilde{\bm{{\mathcal{R}}}}^{-1} \tilde{\mathbf{C}}]^{-1} \tilde{\mathbf{a}}.
\end{equation}
Each beamformer $\tilde{\mathbf{b}}_{nm}$ can be obtained from $\tilde{\mathbf{b}}$ directly, and then the estimated desired SH coefficient $\hat{\tilde{{u}}}_{nm}$ is obtained as
\begin{equation}
  \label{equ:bf_sh_singlestage}
  \hat{\tilde{{u}}}_{nm} = \tilde{\mathbf{b}}_{nm}^H \tilde{\mathbf{x}}.
\end{equation} 

Following \cite{zhang2025spatial}, the estimated desired soundfield pressures over the ROI is obtained by applying the Inverse Spherical Harmonic Transform (ISHT) \cite{rafaely2015fundamentals, williams1999fourier} to the estimated desired SH coefficients $\hat{\tilde{\mathbf{u}}}$.

\section{Simulation and Results}
\label{sec:simulation}
In this section, we evaluate the performance of the proposed direction-preserving SASE algorithm in comparison with an existing SASE algorithm \cite{zhang2025spatial}.
Building on prior work \cite{herzog2020direction, cornelis2009theoretical}, we also implement a modified version of the algorithm in \cite{zhang2025spatial} as an additional baseline.
This variant, referred to as the modified SASE algorithm, preserves the spatial cues of the residual interference soundfield by partially adding original mixed SH coefficients $\tilde{\mathbf{x}}$ to the processed output of the existing SASE algorithm in \cite{zhang2025spatial}.

\subsection{Setup}
\label{ssec:setup}
Two speech signals from LibriSpeech \cite{panayotov2015librispeech} are used as the signals of the desired and interference sources, located at $(1.50, 3.20, 1.60)$ m and $(2.50, 4.20, 1.80)$ m, respectively.
A 32-microphone open spherical array is placed at $(1.50, 4.20, 1.60)$~m, which is the center of the ROI.
The radii of both the microphone array and the ROI are set to $0.042$~m.
The signal-interference-ratio (SIR) is set to $0$~dB, where the 'signal' and 'interference' correspond to the desired and interference source signals measured at their respective source positions, respectively.
A diffuse noise field is generated following \cite{habets2007generating} with a signal-diffuse noise-ratio (SDNR) of $20$~dB.
In addition, white Gaussian noise is added to each microphone as sensor noise, resulting in a signal-sensor noise-ratio (SSNR) of $40$~dB.
The 'signal' in both SDNR and SSNR also refers to the desired source signal.
The acoustic environment is modeled as a room of size $5 \times 6 \times 4$~m with a reverberation time of $T_{60}=0.2$~s.
Room impulse responses (RIRs) are simulated using the image source method \cite{allen1979image, jarrett2012rigid}, with the speed of sound set to $343$~m/s.
The simulated microphone signals are sampled at $16,000$~Hz.

Using the Short-Time Fourier Transform (STFT), the time-domain microphone signals are transformed into the STFT domain with a frame length of $4096$ samples and an overlap of $3072$ samples.
The SHT is then applied to each frequency bin of the resulting STFT-domain microphone signals, with the maximum order set to $N_k$.

Accurate ReHCs of both the desired and interference soundfields are assumed to be available as prior knowledge and are generated following \cite{hu2020semi, markovich2015performance}.
In addition, the SH-domain PSD matrix of interference and noise is estimated from a $5$-frame segment (approximately $0.5$~s) of microphone recordings.
The microphone recordings are captured when the desired source is inactive, such that only interference, diffuse noise, and sensor noise are present.

Following \cite{zhang2025spatial}, the performance of the proposed algorithm is evaluated over the bandwidth ranging from $300$~Hz to $3400$~Hz, where most speech information exists \cite{hioki1998telecommunications, pulakka2013development}.

\subsection{Evaluation}
\label{ssec:eval}

\begin{figure}[t]
\centering
\includegraphics[width=12cm]{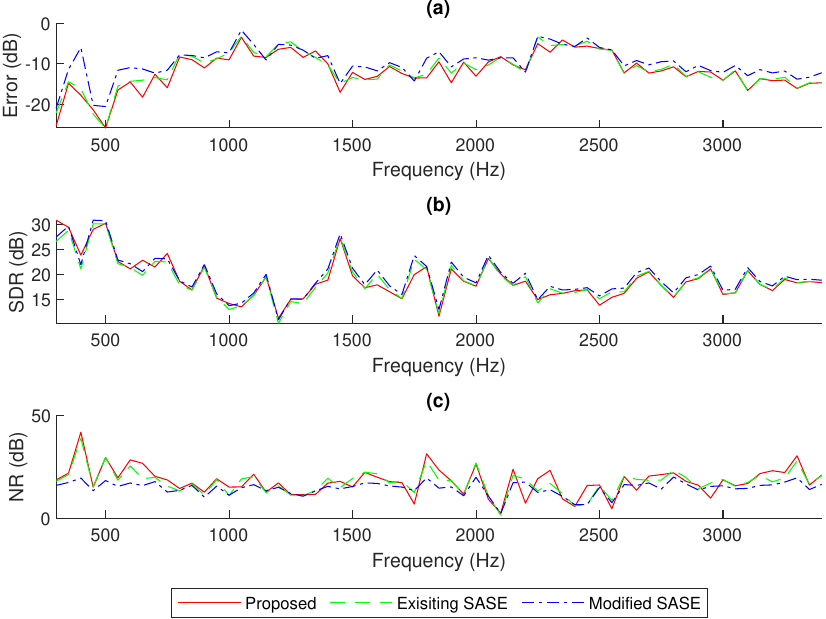}
\caption{The average of Error (a), SDR (b), and NR (c) of the proposed algorithm (red), the existing SASE algorithm (green), and the modified SASE algorithm (blue) inside the spherical ROI.}
\label{fig:SASE}
\end{figure}
The performance of the proposed algorithm is evaluated in terms of estimation error (Error), speech-distortion-ratio (SDR), and noise reduction (NR) across frequencies.
These metrics are averaged over $343$ observation points uniformly distributed within the spherical ROI over $200$ time frames.
The Error, SDR, and NR at these observation points are defined as \cite{herzog2020direction}
\begin{equation}
  \label{equ:error_ob}
  \mathrm{Error}(k) = 10 \log_{10} \frac{\Vert \mathbf{u}_{\text{ob}}(k) - \hat{{\mathbf{u}}}_{\text{ob}}(k) \Vert^2}{\Vert \mathbf{u}_{\text{ob}}(k) \Vert^2},
\end{equation}
\begin{equation}
  \label{equ:sdr}
  \mathrm{SDR}(k) = 10 \log_{10} \frac{\Vert \mathbf{u}_{\text{ob}}(k) \Vert^2}{\Vert \mathbf{u}^\mathrm{res}_{\text{ob}}(k) - \mathbf{u}_{\text{ob}}(k) \Vert^2},
\end{equation}
\begin{equation}
  \label{equ:nr}
  \mathrm{NR}(k) = 10 \log_{10} \frac{\Vert \mathbf{v}^\mathrm{res}_{\text{ob}}(k) + \mathbf{z}^\mathrm{res}_{\text{ob}}(k) + \mathbf{c}^\mathrm{res}_{\text{ob}}(k) \Vert^2}{\Vert \mathbf{v}_{\text{ob}}(k) + \mathbf{z}_{\text{ob}}(k) \Vert^2},
\end{equation}
where $\mathbf{u}_{\text{ob}}(k)$, $\mathbf{v}_{\text{ob}}(k)$, $\mathbf{z}_{\text{ob}}(k)$, $\hat{\mathbf{u}}_{\text{ob}}(k)$ are true desired pressures, true interference pressures, true diffuse noise pressures, and estimated desired pressures, respectively, $\mathbf{u}^\mathrm{res}_{\text{ob}}(k)$, $\mathbf{v}^\mathrm{res}_{\text{ob}}(k)$, $\mathbf{z}^\mathrm{res}_{\text{ob}}(k)$ are residual desired pressures, residual interference pressures, residual diffuse noise pressures, respectively, $\mathbf{c}^\mathrm{res}_{\text{ob}}(k)$ represents distortions introduced by sensor noises, and $\Vert \quad \Vert$ denotes the $2$-norm operation.

As shown in Fig.~\ref{fig:SASE}, the proposed algorithm achieves performance comparable to the existing SASE algorithm in terms of Error, SDR, and NR.
By contrast, the modified SASE algorithm yields the lowest NR across frequency bins, resulting in a deteriorated Error.

To further assess robustness, we evaluate the performance of the three algorithms averaged over the bandwidth under different diffuse noise conditions, as summarized in Table~\ref{tab:metric_vary_snr}.
Here, we vary the SDNR level from $10$~{dB} to $30$~{dB}.
Table~\ref{tab:metric_vary_snr} shows that the proposed algorithm achieves the best average performance in terms of Error and NR under different diffuse noise conditions, while maintaining SDR comparable to the two compared algorithms.
This improvement is likely due to the enhanced interference reduction enabled by incorporating ReHCs of the interference soundfield, similar to \cite{hadad2015theoretical}.
By contrast, the modified SASE algorithm consistently yields the lowest NR and the highest Error across different SDNR levels, despite achieving superior SDR compared to the other algorithms.

\begin{table*}[tb]
	\centering
	\caption{The performance comparison across $200$ time frames against different SDNR levels:  (a) the proposed algorithm, (b) the existing SASE algorithm, and (c) the modified SASE algorithm.}
	\label{tab:metric_vary_snr}  
	\begin{tabular}{|c|c|c|c|c|c|c|c|c|c|} 
	    \hline  
		\multirow{2}{*}{} &  \multicolumn{3}{c|}{SDNR = $10$~{dB}}  & \multicolumn{3}{c|}{SDNR = $20$~{dB}} &  \multicolumn{3}{c|}{SDNR = $30$~{dB}}  \\  
        \cline{2-10}
         & (a) & (b) & (c) & (a) & (b) & (c) & (a) & (b) & (c) \\  
		\hline  
		Error & $-7.1$ & $-6.1$ & $-5.0$ & $-11.8$ & $-11.3$ & $-9.8$ & $-14.2$ & $-14.3$ & $-12.0$ \\
		\hline
        SDR & $19.1$ & $19.1$ & $19.7$ & $19.0$ & $19.0$ & $19.6$ & $18.7$ & $18.9$ & $19.5$ \\
        \hline
        NR & $15.5$ & $14.4$ & $12.9$ & $18.1$ & $17.6$ & $14.8$ & $21.5$ & $21.6$ & $16.3$ \\
        \hline
	\end{tabular}
\end{table*}

To evaluate how well the spatial distributions of the desired soundfield and the residual interference soundfield are preserved, two metrics named the similarity of the desired soundfield $\sigma_{\mathbf{u}, \mathbf{u}_{\mathrm{res}}}$ and the similarity of the interference soundfield $\sigma_{\mathbf{v}, \mathbf{v}_{\mathrm{res}}}$ are presented, respectively, in \cite{herzog2020direction} by
\begin{equation}
\label{equ:similarity_desired}
    \sigma_{\mathbf{u}, \mathbf{u}_{\mathrm{res}}} = \frac{2}{\pi} \sin^{-1} \left( 
\frac{\mathrm{Tr} \left\{ \mathbf{R}_{\mathbf{u}} \mathbf{R}_{\mathbf{u}_{\mathrm{res}}} \right\}}{
\sqrt{ \mathrm{Tr} \left\{ \mathbf{R}_{\mathbf{u}}^2 \right\} } 
\sqrt{ \mathrm{Tr} \left\{ \mathbf{R}_{\mathbf{u}_{\mathrm{res}}}^2 \right\} }}
\right), 
\end{equation}
\begin{equation}
\label{equ:similarity_interference}
    \sigma_{\mathbf{v}, \mathbf{v}_{\mathrm{res}}} = \frac{2}{\pi} \sin^{-1} \left( 
\frac{\mathrm{Tr} \left\{ \mathbf{R}_{\mathbf{v}} \mathbf{R}_{\mathbf{v}_{\mathrm{res}}} \right\}}{
\sqrt{ \mathrm{Tr} \left\{ \mathbf{R}_{\mathbf{v}}^2 \right\} } 
\sqrt{ \mathrm{Tr} \left\{ \mathbf{R}_{\mathbf{v}_{\mathrm{res}}}^2 \right\} }}
\right),
\end{equation}
where $\mathbf{R}_{\mathbf{u}}$, $\mathbf{R}_{\mathbf{v}}$, $\mathbf{R}_{\mathbf{u}_{\mathrm{res}}}$, and $\mathbf{R}_{\mathbf{v}_{\mathrm{res}}}$ are spectro-temporal averaging PSD matrices of true desired soundfield pressures, true interference soundfield pressures, residual desired soundfield pressures, and residual interference soundfield pressures, respectively, and $\mathrm{Tr} \{ \quad \}$ represents the trace operation.

We analyze the similarity achieved by the three algorithms under different diffuse noise conditions ($10$~dB, $20$~dB, and $30$~dB), as summarized in Table~\ref{tab:similarity_vary_snr}.
Table~\ref{tab:similarity_vary_snr} shows that all algorithms achieve a high similarity of the desired soundfield at $0.99$ against different SDNR levels, indicating that the spatial cues of the desired soundfield are well preserved.
By contrast, the existing SASE algorithm yields the lowest similarity of the interference soundfield at about $0.70$ against different SDNR levels, suggesting a noticeable distortion in the spatial cues of the residual interference soundfield.
When the SDNR is $10$~dB, the proposed algorithm achieves the similarity of the interference soundfield at $0.85$, comparable to the modified SASE algorithm.
As the SDNR level increases to $30$~dB, the similarity of the interference soundfield achieved by the proposed algorithm decreases to $0.75$, lower than that of the modified SASE algorithm.
Nevertheless, it remains higher than that of the existing SASE algorithm.

\begin{table*}[tb]
	\centering
	\caption{The similarity analysis across $200$ time frames against different SDNR levels:  (a) the proposed algorithm, (b) the existing SASE algorithm, and (c) the modified SASE algorithm.}
	\label{tab:similarity_vary_snr}  
	\begin{tabular}{|c|c|c|c|c|c|c|c|c|c|} 
	    \hline  
		\multirow{2}{*}{} &  \multicolumn{3}{c|}{SDNR = $10$~{dB}}  & \multicolumn{3}{c|}{SDNR = $20$~{dB}} &  \multicolumn{3}{c|}{SDNR = $30$~{dB}}  \\  
        \cline{2-10}
         & (a) & (b) & (c) & (a) & (b) & (c) & (a) & (b) & (c) \\  
		\hline  
        $\sigma_{\mathbf{u}, \mathbf{u}_{\mathrm{res}}}$ & $0.99$ & $0.99$ & $0.99$ & $0.99$ & $0.99$ & $0.99$ & $0.99$ & $0.99$ & $0.99$\\  
        \hline
        $\sigma_{\mathbf{v}, \mathbf{v}_{\mathrm{res}}}$ & $0.85$ & $0.71$ & $0.89$ & $0.76$ & $0.70$ & $0.97$ & $0.75$ & $0.70$ & $0.98$\\
        \hline	
	\end{tabular}
\end{table*}

\section{Conclusion}
\label{sec:conclusion}
This paper presents a direction-preserving spatial audio signal enhancement (SASE) algorithm in the spherical harmonic domain, designed to capture the desired soundfield and preserve the spatial cues of the desired soundfield and the residual interference soundfield.
Simulation results demonstrate that the proposed algorithm achieves performance comparable to the existing SASE algorithm in terms of estimation error, speech-distortion-ratio, and noise reduction.
Importantly, it better preserves the spatial cues of the residual interference soundfield.
In the future, we plan to design a specific metric to further evaluate the performance of direction-preserving SASE.

\bibliographystyle{unsrt}
\bibliography{sample} 

\end{document}